\documentclass[book,fleqn,10pt]{wlscirep}
\usepackage[utf8]{inputenc}
\usepackage[T1]{fontenc}
\usepackage{xcolor}
\usepackage{amsmath,amsfonts,amsthm,bm} % Math 
\usepackage{changepage}
\usepackage{textcomp,marvosym}
\usepackage{cite}
\usepackage{nameref,hyperref}
\usepackage[right]{lineno}
\usepackage{array}
\usepackage{lastpage,fancyhdr,graphicx}
\usepackage{epstopdf}
\usepackage[aboveskip=1pt,labelfont=bf,labelsep=period,justification=raggedright,singlelinecheck=off]{caption}

\title{A Novel Interpretable Fusion Analytic Framework for Investigating Functional Brain Connectivity Differences in Cognitive Impairments}

\author[1,*]{Yeseul Jeon}
\author[2,*]{Jeong-Jae Kim}
\author[3]{SuMin Yu}
\author[2]{Junggu Choi}
\author[2,4,+]{Sanghoon Han}
\affil[1]{Department of Statistics and Data Science, Yonsei University, Seoul, Republic of Korea}
\affil[2]{Graduate Program in Cognitive Science, Yonsei University, Seoul, Republic of Korea}
\affil[3]{Department of Psychology and Neuroscience, Duke University, Durham, North Carolina, USA}
\affil[4]{Department of Psychology, Yonsei University, Seoul, Republic of Korea}

\affil[*]{These authors contributed equally to this work}

\affil[+]{Corresponding author : sanghoon.han@yonsei.ac.kr}

\begin{abstract}
Functional magnetic resonance imaging (fMRI) data is characterized by its complexity and high--dimensionality, encompassing signals from various regions of interests (ROIs) that exhibit intricate correlations. Analyzing fMRI data directly proves challenging due to its intricate structure. Nevertheless, ROIs convey crucial information about brain activities through their connections, offering insights into distinctive brain activity characteristics between different groups. To address this, we propose a cutting-edge interpretable fusion analytic framework that facilitates the identification and understanding of ROI connectivity disparities between two groups, thereby revealing their unique features. Our novel approach encompasses three key steps. Firstly, we construct ROI functional connectivity networks (FCNs) to effectively manage fMRI data. Secondly, employing the FCNs, we utilize a self--attention deep learning model for binary classification, generating an attention distribution that encodes group differences. Lastly, we employ a latent space item-response model to extract group representative ROI features, visualizing these features on the group summary FCNs. We validate the effectiveness of our framework by analyzing four types of cognitive impairments, showcasing its capability to identify significant ROIs contributing to the differences between the two disease groups. This novel interpretable fusion analytic framework holds immense potential for advancing our understanding of cognitive impairments and could pave the way for more targeted therapeutic interventions.
\end{abstract}

\begin{document}

\flushbottom
\maketitle
\thispagestyle{empty}

\section*{Introduction}

The nature of functional magnetic resonance imaging (fMRI) data, particularly resting--state fMRI, is characterized by its inherent complexity and high dimensionality, forming a correlated matrix that includes signals from brain regions of interest (ROIs) measured at each time point. Several attempts have been made to analyze fMRI data to understand the roles of ROIs in specific tasks or symptoms~\cite{santana2022rs,wang2016abnormal}. Comparing ROIs with fMRI data from different tasks has been one approach to comprehending their mechanisms and identifying differences between groups~\cite{li2018brain,lee2020prediction}. However, interpreting which features of ROIs connections differentiate between two different groups has proven challenging for previous studies. Two main reasons contribute to this difficulty: first, the high--dimensional and correlated structure of fMRI datasets makes it challenging to apply standard statistical models, which rely on the assumption of independent and identically distributed data. In fMRI data, complex interactions and dependencies among ROIs render this independence assumption unrealistic, leading to potentially biased or inaccurate interpretations. Second, identifying group representative features of ROI connections in fMRI data is hindered by the presence of noise caused by individual effects. Each fMRI data unit corresponds to an independent subject, and inherent variability and noise in individual data may obscure the true underlying patterns that differentiate different groups or conditions.

To overcome these limitations, we propose a novel analytic framework that combines deep learning-based classification and statistical modeling while providing visual interpretation through ROIs functional connectivity networks (FCNs) to offer intuitive insights. Deep learning models are well-suits are ed for handling high--dimensional correlated structured data~\cite{du2022eeg}, and we employ self--attention mechanism~\cite{vaswani2017attention} for binary classification which can handle correlated structure data and train their adjacency connections well~\cite{chen2018class,zheng2019stock,sun2020pointgrow} Therefore, we can effectively capturing intricate connectivity patterns among ROIs in fMRI data. The self--attention mechanism focuses on specific input values, leading to improved network information for both local and global connections, thus enhancing prediction accuracy and producing ROIs' attention distributions for each subject. The attention distribution of the ROIs indicates how the self--attention deep learning model trains the correlated structured input data; each row in the attention distribution defines the likelihood of how one specific ROI relates to other ROIs. If the accuracy is sufficient, the output of ROIs attention distribution of each subject is a reliable source to decipher what ROIs connections distinguish the different groups. However, manually comparing these distributions to understand which ROI connections differentiate between groups remains a challenge. To address this challenge, we analyze the ROIs attention distribution using the latent space item-response model (LSIRM)~\cite{jeon2021network}, a statistical network model. We interpret the attention distribution as an item-response matrix~\cite{embretson2013item}, where ROIs represent items, and subjects represent respondents. The LSIRM estimates relationships between respondents and ROIs by modeling the probabilities of positive responses (connections), selecting group representative ROIs with commonly reacted connections within each group. These distinctive features of ROIs connections are visualized on the group summary FCN.

Our framework comprises three key steps. First, we construct FCNs for individual subjects' ROIs by connecting them based on embedded positions using mapper~\cite{chazal2017introduction}. These latent positions, obtained through dimension reduction methods from fMRI data, create individual FCNs, which are then summarized to create group representative FCNs for each group. Despite showing the overall connectivity structure of fMRI data, it remains challenging to identify significant ROI connections differentiating one group from others. As a second step, we perform binary classification based on subjects' FCNs using a self--attention deep learning model~\cite{vaswani2017attention}. To validate the feasibility of our proposed analytic framework, we apply it to classify resting state--fMRI (rs--fMRI) data for different stages of neurodegenerative diseases with varying cognitive impairments. Using resting brain scans from the Alzheimer's Disease Neuroimaging Initiative (ADNI) database, a multisite longitudinal study extensively utilized in biomarker exploration for Alzheimer's disease diagnosis~\cite{jack2008alzheimer, mueller2005ways}, we aim to discover latent functional ROI groups and compare our results with previous findings from over a thousand ADNI publications.

\section*{Results}

\begin{figure}[ht!]
\centering
\includegraphics[scale=0.80]{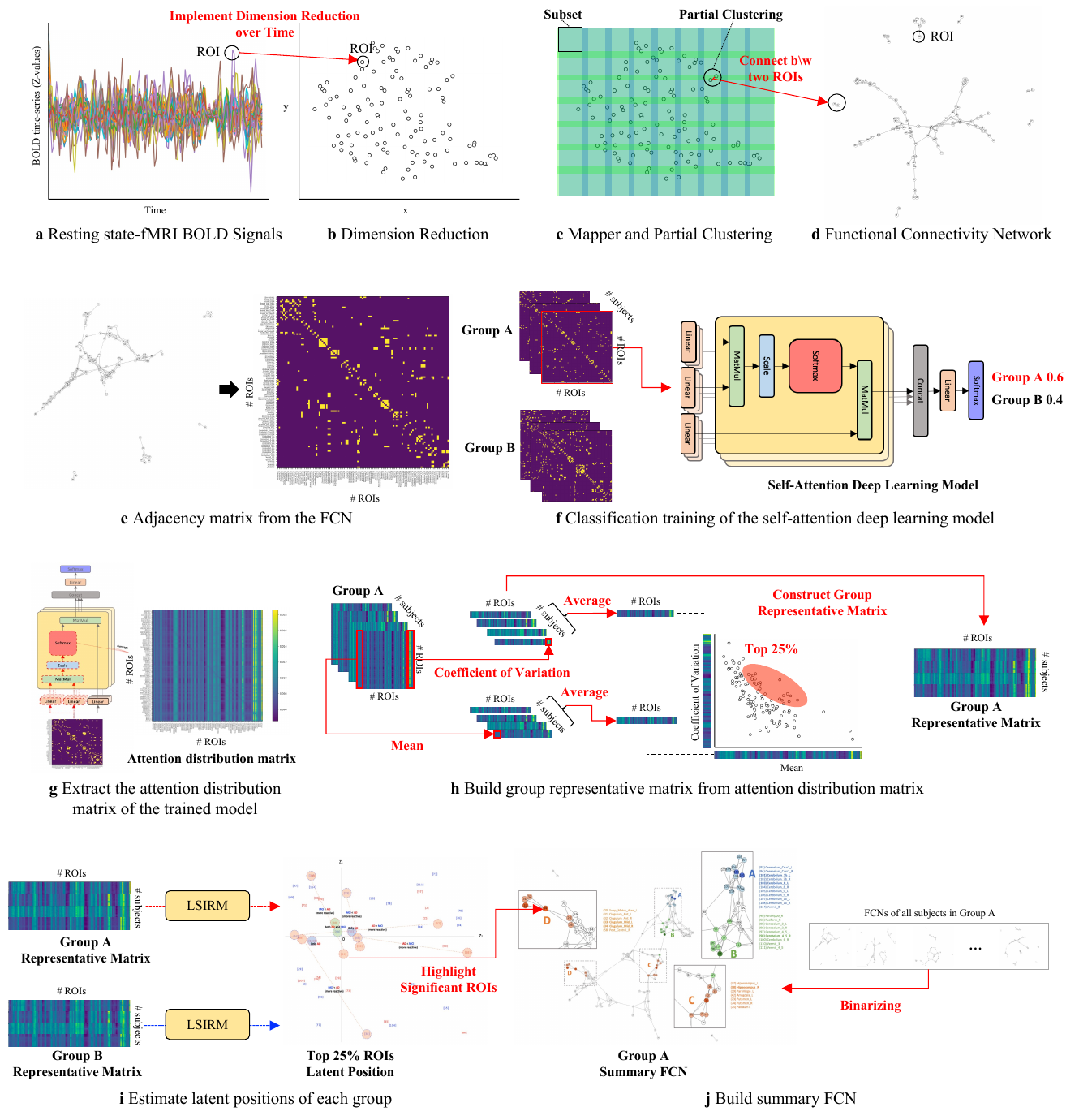}%0.9
\caption{Consider $G$ number of groups and each group has $N_{G}$ number of subjects. {\bf \textbf{Step 1:}} Construct Functional Connectivity Network (FCN) (\textbf{d}) from resting state--fMRI BOLD signals (\textbf{a}) using five different dimension reduction methods: (1) Pearson's r score, (2) Fisher's z score, (3) PCA, (4) t-SNE, and (5) UMAP (\textbf{b}). Estimate the connection among ROIs using partial clustering in Mapper (\textbf{c}). \textbf{Step 2:} Construct adjacency matrix $\mathbf{M}_{g,i}$ for $g=1,\cdots,G$ and $i=1,\cdots,N_g$ from FCN (\textbf{e}). Implement self--attention deep learning model for binary classification (\textbf{f}) for two pairs of diseases and extract the attention distribution matrix $\mathbf{A}_{g,i}$  (\textbf{g}). From the attention distribution matrix of each subject, select meaningful top 25\% ROIs using coefficient of variation and mean of distribution value and construct group representative matrix $\mathbf{X}_{h|g,h}$ (\textbf{h}). \textbf{Step 3:} Using the latent space item--response model (LSIRM), estimate latent positions of each group using the group representative matrix (\textbf{i}). Highlight significant ROIs from result of LSIRM on summary FCN of each group (\textbf{j}).}
\label{fig:proposal}
\end{figure}

In this study, we applied our analysis framework to identify the specific ROIs that differentiate between two pairs of diseases: Alzheimer's Disease (AD) vs. Mild Cognitive Impairment (MCI), AD vs. Early MCI (EMCI), AD vs. Late MCI (LMCI), and EMCI vs. LMCI. We utilized resting-state fMRI (rs--fMRI) data collected from AD, EMCI, MCI, and LMCI from Alzheimer’s disease neuroimaging initiative (ADNI) dataset.

\subsection*{Step1: Functional connectivity networks of each group}

First, we constructed a FCN among brain regions based on their rs--fMRI Blood--Oxygen--Level--Dependent (BOLD) signals. We utilized the automated anatomical labeling (AAL)--116 template to extract 116 rs--fMRI BOLD signals, representing different brain regions. Supplementary Table~\hyperref[supp]{1} in supporting information provides detailed information about the AAL--116 template. Due to the high--dimensional and correlation structure of fMRI data, we implemented dimension reduction over time to embed the high-dimensional correlated structure dataset into low two--dimensional space (Fig.~\hyperref[fig:proposal]{1b}).

To address the subjectivity in determining relevance among regions of interest (ROIs), we adopted mapper~\cite{chazal2017introduction}, a partial clustering method, to identify significant connections between ROIs (represented as Fig.~\hyperref[fig:proposal]{1c}). By assigning ROIs to the same cluster, we considered them connected. Subsequently, we generated FCNs for each set of embedded ROIs obtained from different dimension reduction methods (represented as Fig.~\hyperref[fig:proposal]{1d}). These FCNs captured relationships and connectivity patterns within the high-dimensional correlated fMRI data, representing the data as a connectivity network.

\begin{figure}[ht!]
\centering
\includegraphics[scale=0.9]{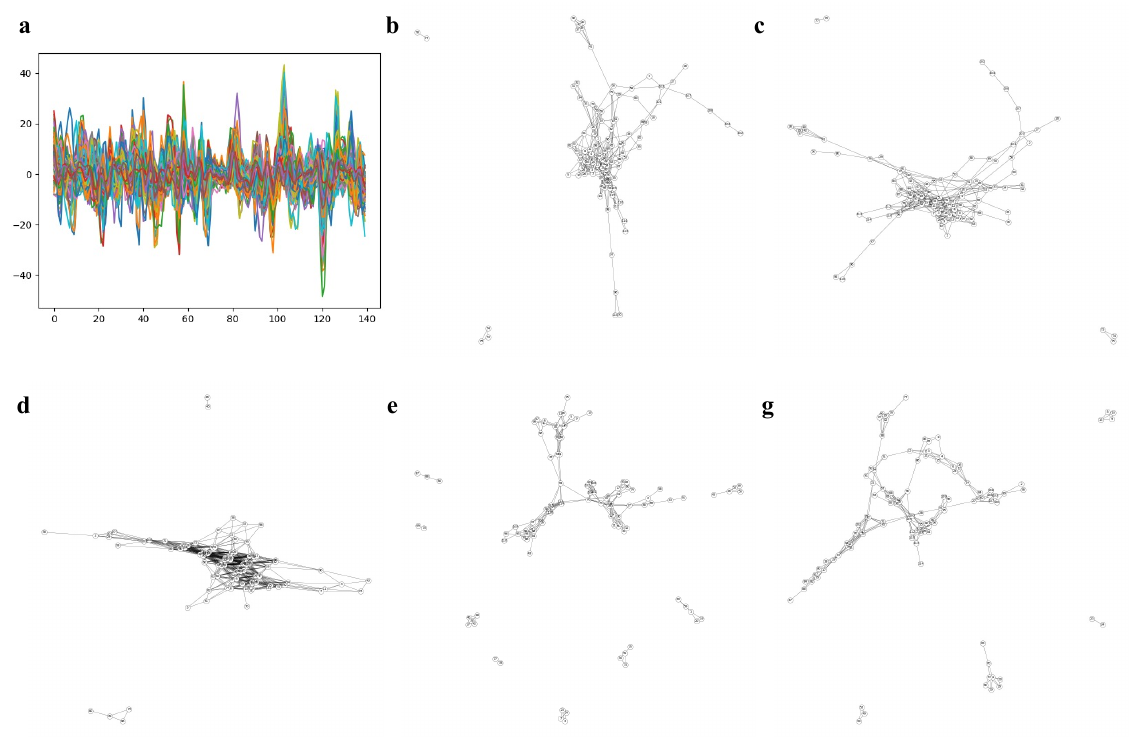}%0.9
\caption{Correlation coefficient based FCNs and dimension reduction based FCNs of AD subject. (\textbf{a}) illustrates the ROIs, rs--fMRI BOLD signals for a sample of subjects with AD. These FCN graphs were generated using various approaches. Firstly, we employed correlation-based methods such as Pearson's r or Fisher's z values (shown in (\textbf{b}) and (\textbf{c}) to establish intricate and interconnected FCNs. However, comprehending the specific attributes of each brain region within these correlation--based FCNs proved challenging. To gain a deeper understanding of the interrelationships between brain regions, we employed dimension reduction techniques to estimate the latent positions of the brain regions. (\textbf{d}) demonstrates the brain region patterns embedded in a 2D space with the highest exploratory power, obtained through PCA in linear space. Additionally, we utilized t--SNE (stochastic space--based FCN) as shown in (\textbf{e}), which assumes that the patterns between brain regions follow a specific probability distribution and learns the degree of similarity between these distributions. Furthermore, we employed UMAP (topological space--based FCN) depicted in (\textbf{f}) to capture the topological similarity of the waveform patterns generated by the ROIs.}
\label{fig:fcn-ad}
\end{figure}

Fig.~\ref{fig:fcn-ad} and Supplementary Figs.~\hyperref[supp]{1}-\hyperref[supp]{3} show each subject's rs--fMRI BOLD signals and two types of FCNs: correlation coefficient-based FCNs and dimension reduction--based FCNs obtained from dimension reduction methods corresponding to MCI, EMCI, and LMCI. By employing multiple approaches (correlation--based and dimension reduction-based), we gained insights into the complex connectivity of ROIs from diverse perspectives, facilitating a comprehensive understanding of their structural characteristics. These FCNs served as inputs for self--attention deep learning model used to classify the two pairs of diseases in our study.

\subsection*{Step2: Attention distribution matrix from self--attention deep learning model}

The FCNs only represented the overall connections within each subject's fMRI data. Our analysis focused on selecting features that differentiate between two pairs of disease groups. To achieve this, we employed self-attention deep learning model. In this model, the input data was the adjacency matrices derived from the FCNs of subjects from different groups (Fig.~\hyperref[fig:proposal]{1e}), and the target was a binary indicator representing the group membership. We applied a total of 128 parallel self--attention and used 116 ROIs. The learning process was a 10--fold cross validation, batch size of 8, dropout of 0.9, Adam~\cite{kingma2014adam} optimizer, learning rate of 0.01 and utilizing the cross entropy loss.

Table~\ref{tab:classification} shows the performance of the self--attention deep learning model. When compared the classification performance to the recent studies~\cite{liu2020multi,wee2019cortical} and the baseline models~\cite{chen2016xgboost,qiu2020development,zunair2020uniformizing} (i.e.,eXtreme Gradient Boosting, Multi Layer Perceptron, Convolutional Neural Networks), our method outperforms in all disease group pairs. Noticeably, the stochastic based and topological based FCN, representing the hidden connectivity among ROIs, yielded the highest accuracy among disease group pairs reflecting the superiority of utilizing high--dimensional dependency ROI structure.

\begin{table}[ht]
\centering
\begin{tabular}{ccccc}
\hline
Method & AD/MCI          & EMCI/AD         & LMCI/AD         & EMCI/LMCI       \\ \hline
Liu, M. et al.~\cite{liu2020multi}                              & 0.8890          & - & - & - \\ 
Wee, C.-Y. et al.~\cite{wee2019cortical} & - & 0.7920          & 0.6520          & 0.6090          \\ \hline
PEARSON+XGBoost                & 0.7949          & 0.8206          & 0.8511          & 0.7802          \\ 
FISHER+XGBoost               & 0.7460          & 0.6254          & 0.6353          & 0.6776          \\ \hline
PEARSON+MLP                & 0.8132          & 0.7600          & 0.7809          & 0.7457          \\ 
FISHER+MLP               & 0.7835          & 0.7533          & 0.7900          & 0.7605          \\ 
LINEAR+MLP      & 0.7461          & 0.7333          & 0.7146          & 0.7452          \\ 
STOCHASTIC+MLP  & 0.7659          & 0.7600 & 0.6427          & 0.7381 \\ 
TOPOLOGICAL+MLP & 0.7819 & 0.7067          & 0.6518 & 0.6838          \\ \hline
PEARSON+CNN                & 0.8654          & 0.7733          & 0.7355          & 0.7667          \\ 
FISHER+CNN               & 0.8648          & 0.7390          & 0.7809          & 0.7267          \\ 
LINEAR+CNN      & 0.7747          & 0.7586          & 0.7891          & 0.7600          \\ 
STOCHASTIC+CNN  & 0.8176          & 0.7657 & 0.7246          & 0.8200 \\ 
TOPOLOGICAL+CNN & 0.8192 & 0.7600          & 0.7155 & 0.7600          \\ \hline
PEARSON+Self--Attn                & 0.8659          & 0.8067          & 0.8809          & 0.8071          \\ 
FISHER+Self--Attn               & 0.8813          & 0.8133          & 0.8718          & 0.8152          \\ 
LINEAR+Self--Attn      & 0.9022          & 0.8467          & 0.8627          & 0.8624          \\ 
STOCHASTIC+Self--Attn  & 0.9033          & \textbf{0.8867} & 0.8900          & \textbf{0.8971} \\ 
TOPOLOGICAL+Self--Attn & \textbf{0.9104} & 0.8733          & \textbf{0.9173} & 0.8695          \\ \hline
\end{tabular}
\caption{The performance of our self--attention deep learning model. When compared to the baseline model, our method outperforms in all disease group pairs.(XGBoost: eXreme Gradient Boosting~\cite{chen2016xgboost}, MLP: Multi Layer Perceptron~\cite{qiu2020development}, CNN: Convolutional Neural Networks~\cite{zunair2020uniformizing}, Self--Attn: Self--attention deep learning model, PEARSON: Pearson’s r--based FCN, FISHER: Fisher’s z--based FCN, LINEAR: Linear space--based FCN, STOCHASTIC: Stochastic space--based FCN, TOPOLOGICAL: Topological space--based FCN)}
\label{tab:classification}
\end{table}

Through the self--attention deep learning model, we obtained the attention distribution (Fig.~\hyperref[fig:proposal]{1f}) by each $i$th subject from group $g$. These attention distributions, denoted as $\mathbf{A}^{(i)}_{(q,r)} \in \mathbb{R}^{116 \times 116 }$, where $i=1,\cdots, N_g$. Here, $N_g$ indicates the number of subjects from each disease group $g=\lbrace AD, MCI, EMCI, LMCI \rbrace$, where $N_{\text{AD}} = 57$, $N_{\text{MCI}} = 78$, $N_{\text{EMCI}} = 93$, and $N_{\text{LMCI}} = 53$. These attention distributions $\mathbf{A}^{(i)}_{(q,r)}$ provide insights into the features that the model focused on when classifying subjects in each disease group against the other pair of group. 

%  We observe that the attention distribution matrix of ROIs remains consistent across the disease groups, with subtle variations in color scheme. This consistency suggests that certain ROIs have consistent associations with other ROIs, indicating their unique characteristics within each disease group.

We regarded this attention distribution $\mathbf{A}^{(i)}_{(q,r)}$ as attention distribution matrix  $\mathbf{Y}^{(i)}_{g|g,h}$, for $g \neq h$ and $g,h= 1,\cdots,G$, where each row and column corresponds to ROIs, and the values indicate the significance of each ROI's contribution to the classify the subject $i$ in group $g$ against group $h$ ~\cite{vig2019multiscale} (Fig.~\hyperref[fig:proposal]{1g}). Although the resting--state data is minimally affected by external factors, the classification accuracy of 90\% demonstrates that the attention distribution matrix of each disease group indeed capture subtle differences. Thus, we can infer that the attention matrices provide valuable information for distinguishing between the two pairs of disease groups.

\begin{figure}[ht!]
\centering
\includegraphics[scale=0.9]{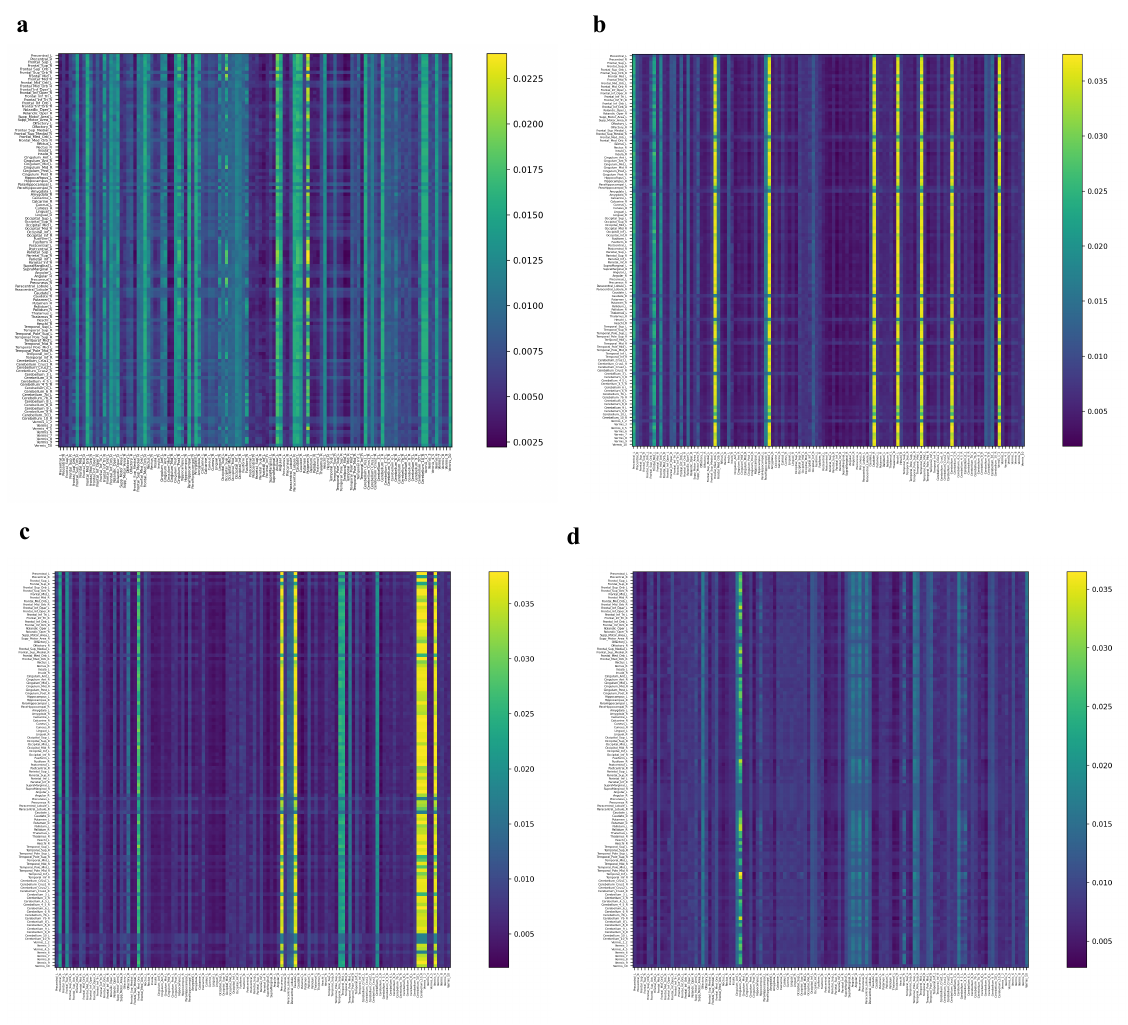}%0.9
\caption{Attention distribution matrices for four subjects in the AD group from the self--attention deep learning model designed for AD and MCI classification, utilizing topological-based FCNs as input. In (\textbf{a}), the attention distribution matrix of a subject highlights high attention values in Putamen\_R, Angular, and Paracentral areas, indicating the self--attention deep learning model's focus on assigning this subject to the AD category over MCI. Similarly, (\textbf{b}) depicts another subject's attention distribution matrix from the AD group, with elevated values in ParaHippocampal\_L, Amygdala\_L, Caudate\_R, Heschl\_L, Temporal\_Mid\_R, Cerebellum\_3\_L, and Vermis\_1\_2. (\textbf{c}) illustrates prominent values in Precuneus\_L, Caudate\_L, Vermis\_6, and numerous regions within the Cerebellum. Lastly, (\textbf{d}) exhibits elevated values in Cingulum\_Ant.}
\label{fig:attention-ad}
\end{figure}

\begin{figure}[ht!]
\centering
\includegraphics[scale=0.9]{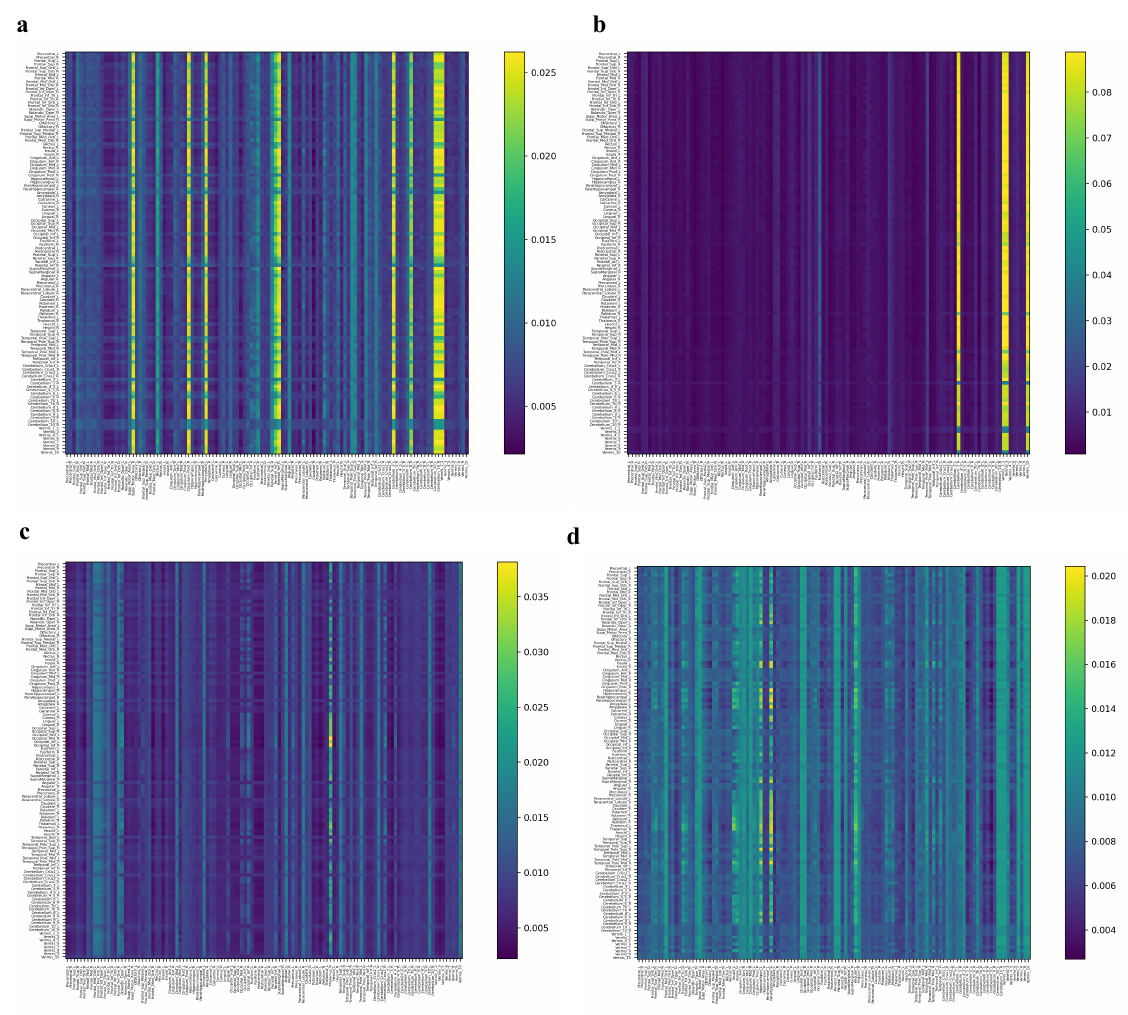}%0.9
\caption{Attention distribution matrices of four subjects within the MCI group, derived from the self--attention deep learning model designed for AD and MCI classification, employing topological--based FCNs as input. In (\textbf{a}), the attention distribution matrix for a subject displays significant attention values in regions like Supp\_Motor\_Area\_R, Cingulum\_Post\_R, Amygdala\_L, Pariental\_Inf, Vermis\_1\_2, and a substantial portion of the Cerebellum. These attention patterns suggest the self--attention deep learning model's emphasis on classifying this subject within the MCI category rather than AD. Similarly, (\textbf{b}) portrays another subject's attention distribution matrix from the MCI group, with elevated values in Fusiform\_R, Cerebellum\_3\_R, and a substantial portion of Vermis. In (\textbf{c}), significant attention values are observed in Thalamus\_R. Lastly, (\textbf{d}) exhibits pronounced attention values in Hippocampus\_L and ParaHippocampal\_R.}
\label{fig:attention-mci}
\end{figure}

Fig.~\ref{fig:attention-ad} and Fig.~~\ref{fig:attention-mci} represent attention matrices for four randomly selected subjects from the AD and MCI groups, respectively. These matrices are the outcomes of the self--attention deep learning model employed for AD and MCI classification, utilizing FCNs derived from topological dimension reduction techniques as inputs. In general, higher attention values assigned to specific ROIs suggest their significance in classifying subjects into respective groups. For instance, in Fig.~\hyperref[fig:attention-ad]{3a}, Putamen\_R shows higher attention value  that highlight a consistent decline in Putamen's volume in the AD group~\cite{de2008strongly}. Similarly, in Fig.~\hyperref[fig:attention-ad]{3b} and Fig.~\hyperref[fig:attention-ad]{3c}, Caudate\_R, Caudate\_L,ParaHippocampal\_L, Cerebellum, Cingulum\_Ant\_L, and Cingulum\_Ant\_R have high values which also show significant ROIs marker in AD group~\cite{kesslak1991quantification,bobinski1999histological,he2007regional,catheline2010distinctive}.  

On the other hand, Fig.~\hyperref[fig:attention-mci]{4a} has high value at Supp\_Motor\_Area\_R and Cingulum\_Post\_R~\cite{lin2014cingulum} which ROIs linked to motor function and exercise~\cite{bai2011mapping,schmahmann2001function,aggarwal2006motor}. Similarly, in Fig.~\hyperref[fig:attention-mci]{4b}, Fig.~\hyperref[fig:attention-mci]{4c}, and Fig.~\hyperref[fig:attention-mci]{4d}, Cerebellum, Vermis~\cite{van2021grey}, Thalamus\_R~\cite{li2013discriminative,cai2015changes}, Hippocampus\_L, and ParaHippocampal\_R~\cite{hamalainen2007increased} emerge as significant ROIs in the MCI group. The combination of these results with the classification accuracy outlined in Table~\ref{tab:classification} underscores the effectiveness of our well--trained self--attention deep learning model in generating meaningful outcomes. Nonetheless, individually interpreting each subject within each group can be time--intensive, and potential noise from individual origins might exist within each self--attention distribution matrix. Consequently, we proceed to extract group representative features employing the latent space item--response model.

To identify the representative features of ROIs connections that differentiate between two groups (e.g., $g$ and $h$), we constructed a group representative matrix $\mathbf{X}{g|g,h} \in \mathbb{R}^{N_g \times 116}$ from each individual attention distribution matrix $\mathbf{Y}^{(i)}{g|g,h}$ for $i=1,\cdots,N_g$, where each row represents a subject and each column represents an ROI (Fig.~\hyperref[fig:proposal]{1b}). The values in the group representative matrix $\mathbf{X}_{g|g,h}$ correspond to the coefficient of variation among each column of the each individual attention distribution matrix $\mathbf{Y}^{(i)}_{g|g,h}$ from group $g$. A high value for a certain ROI in $\mathbf{X}_{g|g,h}$ indicates that this ROI shows a unique pattern in the corresponding individual attention distribution matrix, contributing to the classification of that individual into a specific group. Additionally, we compiled a list of the ROIs of interest that meet the criteria of being in the top 25\% in terms of both high coefficient of variation and high mean values in the attention distribution matrices $\mathbf{Y}^{(i)}_{g|g,h}$ from group $g$ (Fig.~\hyperref[fig:proposal]{1h}). These top 25\% ROIs with high mean values imply frequent interactions with other ROIs, while those with high coefficient of variation imply that those ROIs have signals showing a non--uniform patterns among subjects. Table~\ref{tab:q1} displays the top 25\% unique ROIs from each group in comparison with another pair of disease groups of AD and MCI. Additional details regarding the top 25\% ROIs for other pairwise group comparisons are available in the Supplementary Table~\hyperref[supp]{2}.

\begin{table}[ht!]
\centering
\begin{tabular}{ccc}
\hline
{Top}    &  AD        & MCI        \\ \hline
Top--1  & Postcentral\_L       & Cingulum\_Mid\_L       \\
Top--2  & Postcentral\_R       & Postcentral\_L         \\
Top--3  & Temporal\_Inf\_L     & Fusiform\_R            \\
Top--4  & Supp\_Motor\_Area\_R & Precentral\_R          \\
Top--5  & Fusiform\_L          & Pallidum\_L            \\
Top--6  & Cingulum\_Mid\_L     & Temporal\_Inf\_L       \\
Top--7  & Fusiform\_R          & Fusiform\_L            \\
Top--8  & Cerebelum\_8\_R     & Supp\_Motor\_Area\_R   \\
Top--9  & Cerebelum\_6\_L     & Insula\_L              \\
Top--10 & Putamen\_L           & Cerebelum\_6\_R       \\
Top--11 & Cerebelum\_8\_L     & Cerebelum\_4\_5\_R    \\
Top--12 & Precentral\_R        & Postcentral\_R        \\
Top--13 & Thalamus\_L          & Cerebelum\_4\_5\_L    \\
Top--14 & Temporal\_Mid\_R     & Cerebelum\_6\_L       \\
Top--15 & Cerebelum\_4\_5\_R  & Putamen\_R             \\
Top--16 & Insula\_R            & Cerebelum\_8\_R       \\
Top--17 & Rolandic\_Oper\_R    & Cerebelum\_Crus2\_R   \\
Top--18 & Precentral\_L        & SupraMarginal\_R       \\
Top--19 & Temporal\_Inf\_R     & Rolandic\_Oper\_R      \\
Top--20 & Insula\_L            & Hippocampus\_R         \\
Top--21 & Putamen\_R           & Cingulum\_Mid\_R     \\
Top--22 & Cerebelum\_7b\_R    & Pallidum\_R            \\
Top--23 & Temporal\_Mid\_L     & Thalamus\_L          \\
Top--24 & Hippocampus\_R       & Putamen\_L             \\
Top--25 & Cerebelum\_4\_5\_L  & Vermis\_4\_5           \\
Top--26 & Cerebelum\_10\_R    & Paracentral\_Lobule\_L \\
Top--27 & Cingulum\_Mid\_R     & Vermis\_8              \\
Top--28 & Pallidum\_L          & Precentral\_L          \\
Top--29 & Cerebelum\_7b\_L    & Supp\_Motor\_Area\_L   \\
\end{tabular}
\begin{tabular}{ccccc}
\end{tabular}
\caption{Top 25\% ROIs that show differences between disease group of AD and MCI.}
\label{tab:q1}
\end{table}

\subsection*{Step3: Group representative features using the latent space item-response model}

In the previous Step2, we obtain the group representative matrices $\mathbf{X}_{h|g,h}$ for $g \neq h$ and $g,h= 1,\cdots,G$. To capture the group representative ROIs features that commonly reacted among subjects (shown in Fig.~\hyperref[fig:proposal]{1i}), we applied LSIRM to each group representative matrix from  pair of group $\mathbf{X}_{h|g,h} \in \mathbb{R}^{N_g \times R}$. We estimated the latent positions of ROIs $\bf{V}= \{{\bf v }_i\}$, for $i=1,\cdots,116$ using MCMC. The MCMC ran 55,000 iterations, and the first 5,000 iterations were discarded as burned-in processes. Then, from the remaining 50,000 iterations, we collected 10,000 samples using a thinning of 5. We used two-dimensional Euclidean space to estimate the latent positions of ROIs. Additionally, we set 0.005 for $\boldsymbol\beta$ jumping rule, 0.005 for $\boldsymbol\theta$ jumping rule, and 0.005 for ${\bf w}_j$ and 0.003 ${\bf z}_i$ jumping rules. Here, we fixed prior $\boldsymbol\beta$ follow $N(0,1)$. We set $a_{\sigma}=b_{\sigma}=0.001$. LSIRM takes each matrix ${\bf X}_g$ as input and provides the ${\bf O}_g$ matrix as output after the Procrustes-matching within the model. Since we calculated topics' distance on the 2-dimensional Euclidean space, ${\bf O}_k$ is of dimension $116 \times 2$. To overcome the identifiable issues from the invariance property, we applied {\tt oblimin} rotation to the estimated topic position matrix $\mathbf{O}_{k\%}^{*}$ using the R package {\tt GPArotation} (\url{https://cran.r-project.org/web/packages/GPArotation/index.html}).

\begin{figure}[ht!]
\centering
\includegraphics[scale=0.7]{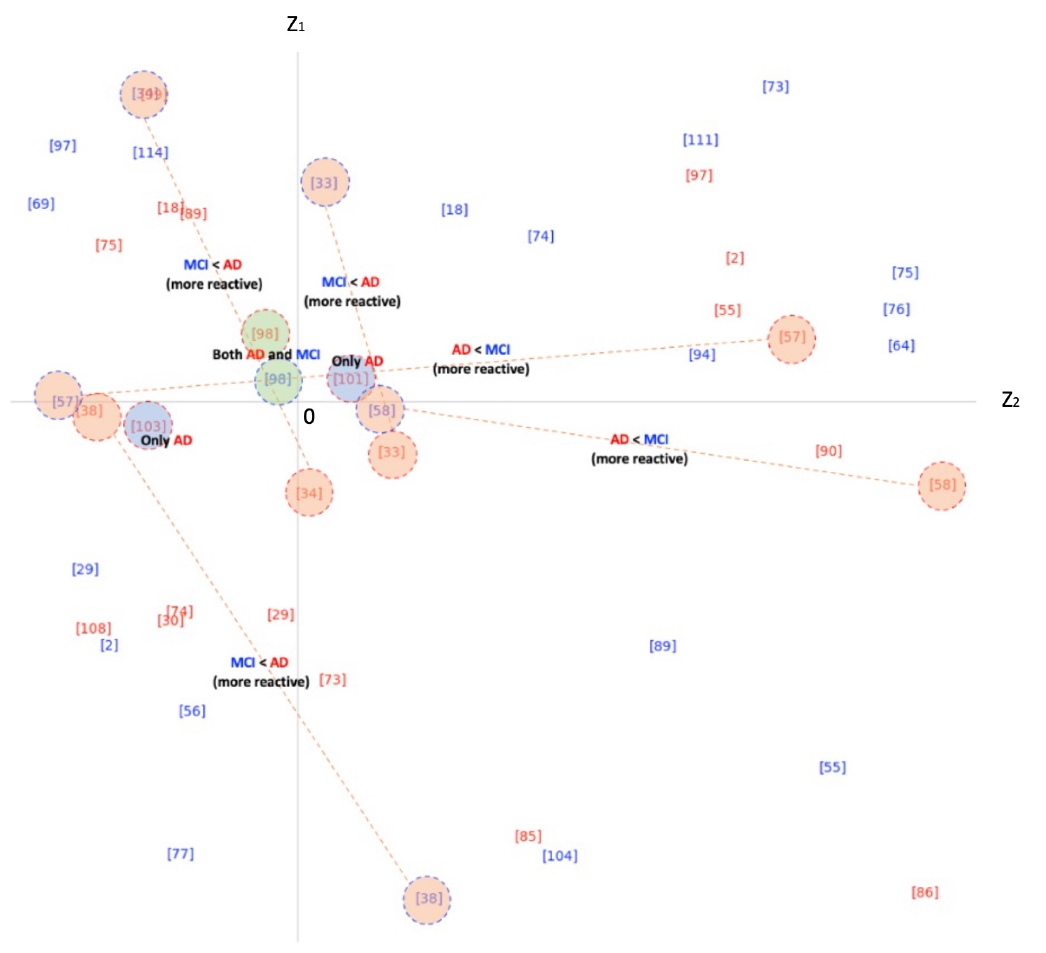}%0.9
\caption{Latent positions of ROIs in 2--dimensional Euclidean space. Red color of numbers indicate Top 25\% ROIs from AD group and blue color of numbers indicate top 25\% ROIs from MCI group. Latent positions located closer to the origin suggest a higher likelihood of common interactions among subjects within the group. There are three scenarios: (1) more reactive pattern, where when comparing two groups, only the latent position of one group is located near the origin while the other group's latent position is situated outside the origin(orange color); (2) both group, where latent positions of both groups are near the origin (green color); and (3) only, indicating that a specific ROI is ranked in the top 25\% within one group (blue color).}
\label{fig:lsrm}
\end{figure}

Based on the estimated latent positions, we successfully identified ROIs that exhibited common reactions within their respective groups. Fig.~\ref{fig:lsrm} exclusively displays the latent positions of the top 25\% ROIs from each group. As depicted in Fig.~\ref{fig:lsrm}, the latent positions of ROIs are visualized in Euclidean space. Through the comparison of ROIs' latent positions between two distinct groups, we were able to pinpoint the ROIs with disparate patterns. In this representation, red-colored numbers signify the Top 25\% ROIs from the AD group, while blue-colored numbers correspond to the top 25\% ROIs from the MCI group. Notably, ROIs positioned closer to the origin in the latent space suggest a heightened likelihood of shared interactions among subjects within the same group. For instance, both the number 98 ROIs from the AD and MCI groups are located close to the origin and marked in green, indicating their significant roles in both AD and MCI. Conversely, the number 101 and 103 ROIs are exclusively part of the AD group's top 25\%. Moreover, other numbers, color highlighted in orange, indicate that only one group of ROIs possesses latent positions situated near the origin. These ROIs can be interpreted as significant features that exhibit meaningful reactions exclusively in comparison to the other group.

In our analysis framework, therefore, we specifically focus on the ROIs that meet two criteria: being ranked in the top 25\% listed up in Step2, and having latent positions that are located near the origin. These criteria indicate that these ROIs exhibit distinct patterns among subjects and can be considered as representative features of each group. This selection process helps us identify the main features that are prominent and generalize well across the groups. We visually highlight these selected ROIs on the summary FCN from each group. As shown in  Fig.~\hyperref[fig:proposal]{1j}, the summary FCN for each group is obtained by averaging the connectivity of each node across all subjects within the group, and a threshold of 0.2 is applied to define the connections.

Fig.~\ref{fig:summnet_ad_mci} and Supplementary Figs.~\hyperref[supp]{4}-\hyperref[supp]{6} show the differences in disease network between the two groups. Blue indicates meaningful regions that show different values from the attention distribution matrix when compared to the other disease groups. The orange, on the other hand, indicates ROIs that were selected before analysis to be meaningful in both disease groups, but were shown to only be meaningful in one group post--analysis. Finally, green indicates regions that were meaningful in both disease groups.

\subsection*{Interpretation of summary FCN from each group}

\begin{figure}[ht!]
\centering
\includegraphics[scale=0.9]{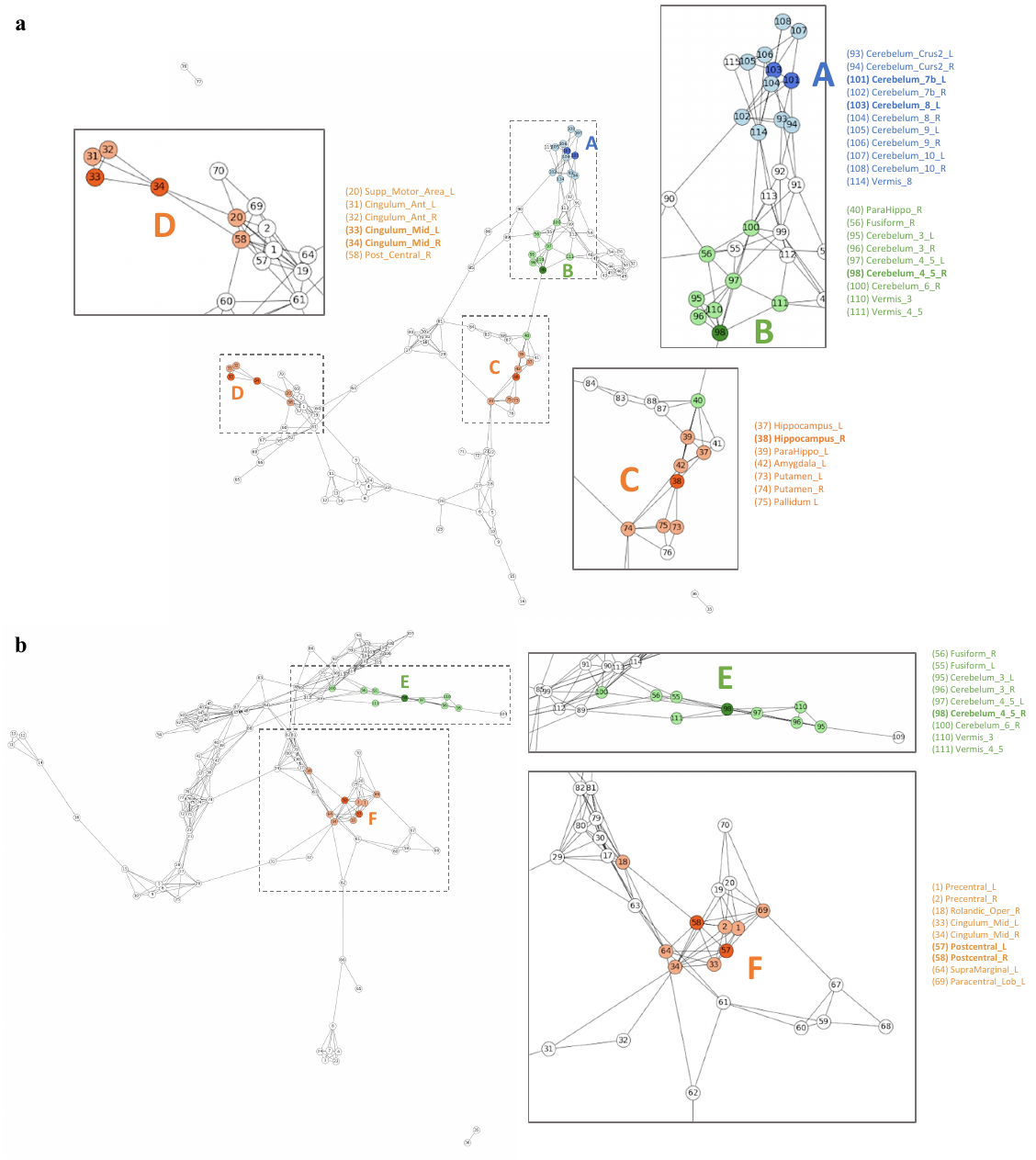}%0.9
\caption{\textbf{(a)} AD group summary FCN and \textbf{(b)} MCI group summary FCN. The darker saturation colors indicates meaningful ROIs of each group using LSIRM and ROI nodes which are directly connected to this meaningful ROIs are represented by the same color with a lower brightness level. Here C, D, and F indicate clusters that exhibit more pronounced responses in the respective group than in the comparative group (orange colored cluster), with B and E representing clusters responsive in both diseases (green colored cluster), while A signifies a cluster responsive solely in AD and not in MCI (blue colored cluster).}
\label{fig:summnet_ad_mci}
\end{figure}

Fig.~\ref{fig:summnet_ad_mci} and Supplementary Figs.~\hyperref[supp]{4}-\hyperref[supp]{6} show the differences in disease network between the two groups. ROIs colored in blue indicate their selection as the top 25\% group from the attention distribution matrix in one group, yet they do not appear as prominently significant in another group. The orange--colored ROIs indicate that they are meaningful only in one group, as revealed by the comparison between latent positions from each group. Finally, the green colored ROIs indicate that they were found to be meaningful in both disease groups. 
Utilizing the property of latent positions estimated from LSIRM~\cite{jeon2021network}, we managed to decode the structural connections among ROIs and identify ROIs that exhibited consistent significance across all subjects within each disease group. In order to see the overall connectivity, we merged the outcomes of LSIRM with FCNs, assigning colors to the significant ROIs and their interconnectedness with other ROIs in FCNs. The higher saturation colors indicates meaningful ROIs features from LSIRM and the ROI nodes which are directly connected to the meaningful ROIs are represented by the same color with a lower brightness level. We can regard this connections as cluster. 

\paragraph{AD/MCI}
\label{sec:result-step3-admci}

According to Fig.~\hyperref[fig:summnet_ad_mci]{6a}, Cluster A is comprised of  Cerebelum\_7\_L (101) and Cerebelum\_8\_L (103). These two regions did not show activity in MCI, and the majority of regions that reacted in AD were connected to the Cerebellum regions. This distinction becomes evident as the AD group exhibits a diminished grey matter volume in the cerebellar anterior lobe in contrast to the non-AD group, as observed in prior research ~\cite{reiman2012brain}. 

Cluster C with Hippocampus\_R (38) and cluster D with Cingulum\_Mid\_L (33) and Cingulum\_Mid\_R (34)  of Fig.~\hyperref[fig:summnet_ad_mci]{6a} show the cluster of regions and their direct connectivity that were more reactive in AD compared to MCI. Hippocampus\_R (38) of cluster C showed greater reactivity in AD when compared to MCI and Hippocampus (37, 38), ParaHippo (39, 40), Putamen (73, 74), Pallidum (75, 76), Amygdala (41, 42) are densely populated in this area. We discovered that the Hippocampus (37, 38) plays an important role in expressing AD characteristics (Fig.~\hyperref[fig:summnet_ad_mci]{6a} Cluster C). Many studies have shown that having  Hippocampus (37, 38) dysfunction affects memory ~\cite{spaniol2009event,small2011pathophysiological,delbeuck2003alzheimer}. Through our methodology, we also identified associations between the Hippocampus (37, 38), Putamen (73, 74), and ParaHippo (39, 40). These regions have been previously associated with cognitive impairment in Alzheimer's disease~\cite{kesslak1991quantification,bobinski1999histological,de2008strongly}.

Cluster F of Fig.~\hyperref[fig:summnet_ad_mci]{6b} shows ROIs, Postcentral\_L (57) and Postcentral\_R (58), that were more reactive in MCI compared to AD. This Postcentral (57, 58) is directly connected to Cingulum\_Mid (33, 34), Precentral (1, 2), Paracentral\_Lob\_L (69). According to our findings, Cingulum\_Mid (33, 34) is linked to the Postcentral (57, 58), Precentral (1, 2), and Paracentral\_Lob (69, 70), all of which are known to process motor information. The fluorodeoxyglucose(FDG) positron emission tomography(PET) modality has been used to investigate all four of the above--mentioned areas as relevant indicators in MCI~\cite{xu2016prediction}.

Cluster B of Fig.~\hyperref[fig:summnet_ad_mci]{6a} and Cluster E of Fig.~\hyperref[fig:summnet_ad_mci]{6b} corresponds to Cerebelum 4\_5\_R (98) that reacted to both AD and MCI. Both results show that Cerebelum 4\_5\_R (98) is not only connected with other Cerebellum regions, but is also directly connected to Fusiform\_L (55) and Fusiform\_R (56), regions that are related to facial recognition~\cite{kanwisher1997fusiform}. According to the global hub node centrality analysis~\cite{zhang2020investigation}, Fusiform (55, 56) and Cerebellum regions play important roles in constituting the key makeup of disease characteristics of MCI.

\paragraph{AD/EMCI}
\label{sec:result-step3-ademci}

Cluster A and B of Supplementary Fig.~\hyperref[supp]{4a}, which are Hippocampus\_L (37), Lingual\_R (48), Cerebelum\_4\_5\_L (97), were found to be meaningful regions not in EMCI but only in AD. Hippocampus (37, 38) in both hemispheres are directly connected. These regions are also directly connected to ParaHippo (39, 40), Putamen (73, 74), Pallidum (75, 76), Amygdala (41, 42) and the results are similar to the results described in Section~\ref{sec:result-step3-admci}. Fusiform\_R (56)and Cerebelum\_8\_L (103) are directly connected to Hippocampus\_R (38), and are similar to cluster A and B of Supplementary Fig.~\hyperref[supp]{4a}. Hippocampus\_L (37) and  Lingual\_R (48) are not only directly connected to Lingual\_L (47), but also to Calcarine (43, 44), Cuneus (45, 46), Fusiform\_L (56) and Cerebelum\_6\_L (99). Cluster D and E of Supplementary Fig.~\hyperref[supp]{4a} were more active in AD relative to EMCI and included the Hippocampus\_R (38), Rolandic\_Oper\_R (18) regions. We are able to see that Rolandic\_Oper\_R (18) is directly connected to Putamen (73, 74), Pallidum (75, 76) and Heschl\_L (79). Cluster F of Supplementary Fig.~\hyperref[supp]{4b} was active in EMCI but not AD, and Cerebelum\_9\_R (106) was analyzed. This region was adjacent to Cerebelum\_Crus2\_R (94), Cerebelum\_7b\_R (102) and Cerebelum\_9\_L (105). Cluster H, I and J of Supplementary Fig.~\hyperref[supp]{4b} are regions that were more active in EMCI relative to AD, and regions Cingulum\_Mid\_L (33), Cingulum\_Mid\_R (34), Pallidum\_L (75) and Cerebelum\_Crus2\_L (93) were analyzed. Cingulum\_Mid is directly connected to Precentral (1, 2), Supp\_Motor (19, 20), Postcentral (57, 58) and Supramarginal (63, 64). These linkages have lately been examined in relation to planning and cognitive control processing ~\cite{domic2021theta,cavanagh2014frontal}.

Cluster C of Supplementary Fig.~\hyperref[supp]{4a} and cluster G of Supplementary Fig.~\hyperref[supp]{4b} are regions that were active in both AD and EMCI, and corresponds to Fusiform\_L (55) and Fusiform\_R (56). Fusiform (55, 56) is directly connected with Hippocampus (37, 38) and ParaHippo (39, 40). Likewise in previous studies~\cite{apostolova2006conversion,li2013discriminative,zhu2009quantitative}, there are connections between Hippocampus (37, 38) and ParaHippo (39, 40), Putamen (73, 74), Pallidum (75, 76), and Amgydala (41, 42).

\paragraph{AD/LMCI}
\label{sec:result-step3-adlmci}

Cluster A of Supplementary Fig.~\hyperref[supp]{5a} was active in AD but not LMCI, and Temporal\_Mid\_R(86) was analyzed. Cluster B of Supplementary Fig.~\hyperref[supp]{5a} reacted more in AD relative to LMCI, and Cerebelum\_6\_L (99) was analyzed. Not only is this region connected with multiple Cerebelum (91, 92, 100) areas, but is also connected to Fusiform\_L (55), Lingual (47, 48), multiple Vermis (112, 113, 114). Cluster C of Supplementary Fig.~\hyperref[supp]{5b} was active only in LMCI and not AD, and Rolandic\_Oper\_R (18) was analyzed. This region was connected with Heschl (79, 80), Insula (29, 30) and Temporal\_Sup (81, 82). Cluster D, E, and F of Supplementary Fig.~\hyperref[supp]{5b} are regions more active in LMCI relative to AD, and regions Putamen\_L (73), Cerebelum\_4\_5\_R (98) and Vermis\_8 (114) were analyzed. Putamen (73, 74) is connected to Olfactory (21, 22), Hippocampus (37, 38), Amygdala (41, 42), Pallidum (75, 76) and Thalamus\_L (77).

\paragraph{EMCI/LMCI}
\label{sec:result-step3-emcilmci}

Cluster A, B, and C of Supplementary Fig.~\hyperref[supp]{6a} were regions that were only active in EMCI, and Frontal\_Inf\_Orb\_R (16), Frontal\_Med\_Orb\_R (26), and Cerebelum\_3\_R (96) were analyzed. Cerebelum\_3\_R (96) is directly connected to Cerebelum\_3\_L (95) and Vermis\_3 (110). ROIs that are connected to Frontal\_Inf\_Orb (15, 16) and Frontal\_Inf\_Tri (13, 14), can be grouped as Frontal regions. We can also see that they are directly connected to Putamen\_R (74). The Frontal\_Med\_Orb\_R (26) is directly connected to Frontal\_Med\_Orb\_L (25), Rectus (27, 28), and Frontal\_Sup\_Orb\_R (6). Cluster D are regions that were more active in EMCI relative to LMCI, and include Putamen\_L (73), Pallidum\_L (75) regions. The ROIs that were primarily connected to these regions can largely be defined as Caudate (71, 72), Pallidum\_R (76), Thalamus (77, 78), Hippocampus (37, 38), and Insula (29, 30). Other regions include Rolandic\_Oper\_L (17), Amygdala\_R (42), Fusiform\_L (55), and Cerebelum\_8\_L (103). Supplementary Fig.~\hyperref[supp]{6b}, on the other hand, shows the FCN extracted for LMCI, which shows no significant ROIs that were significantly active only in LMCI. There is, however, cluster E, that shows ROIs more active in LMCI relative to EMCI. This cluster is comprised of ROIs connected to Temporal\_Mid\_R(86) and Cerebelum\_6\_L (99).  Temporal\_Mid\_L (85) and Temporal\_lnf\_R(90) are ROIs connected with Temporal\_Mid\_R (86). ROIs connected with Cerebelum\_6\_L (99) are largely Fusiform (55, 56), Lingual (47, 48), and multiple Vermis (112, 113). Other regions include Cerebelum\_Crus1\_L (91), Cerebelum\_4\_5\_L (97), and Cerebelum\_6\_R (100).

% \subsection*{Scientific findings}

\section*{Discussion}
Despite the wealth of information contained within fMRI data, our study introduces a pioneering analytical framework that offers an interpretable and enlightening approach to investigating disparities in connections among regions of interest (ROIs) within two distinct pairs of cognitive impairment groups. By effectively addressing the intricate challenges inherent in fMRI data analysis, our proposed methodology yields significant and meaningful  connections of ROIs.

Our fundamental concept centers around a meticulous examination of the distinctive attributes within regions of interest (ROIs) networks as contrasted with other medical conditions. This scrutiny is achieved through a dissection of the results emanating from a classification model founded on self-attention deep learning. In an effort to delve into the outcomes of this self-attention deep learning model, we employ a statistical network model called LSIRM, which is adept at managing correlated structured data and affords an intuitive interpretation of the outcomes. Furthermore, as a foundational step, we construct a functional connectivity network (FCN) that offers a visual representation of the interconnections among ROIs. This FCN serves as a means to dissect the intricate patterns embedded within the high-dimensional and correlated fMRI data. 

This network facilitates a succinct depiction of the overall structure by mapping complex data into a lower-dimensional arrangement, enabling the elucidation of relationships between the functions of different ROIs. By employing each subject's FCN, we gauge the distribution of attention among ROIs through a self-attention deep learning model. Through this novel approach, classification accuracy between two sets of different disease groups markedly improves in comparison to prior research efforts. Consequently, the distribution of attention among ROIs sufficiently reveals concealed mechanisms that differentiate various disease groups. Nonetheless, understanding the implications of this distribution of attention is not inherently intuitive. Moreover, the attention model yields an individual distribution for each subject's ROIs. To gain insight into the mechanisms underlying disease distinctions, it becomes imperative to synthesize these distinctions comprehensively. Addressing this need, we analyze the matrix of attention distributions among ROIs, denoted as $\mathbf{A}$, using a latent space item-response model. Through the modeling of interactions among ROIs, the estimated latent positions of these regions offer intuitive information about the ROIs that commonly elicit responses among subjects within each disease group. Building upon these selected ROIs, we emphasize distinctive ROIs within summary FCNs for each disease group, thereby revealing deeper insights into the nuances of various conditions. Furthermore, we delineate subgroups of connections within summary FCNs for each disease group, thereby facilitating a more profound understanding of the intricacies inherent in distinct illnesses. Our methodology has also unearthed significant biological insights, which have been consistently validated across multiple studies. Our research not only yields results that align with extensive prior studies but also identifies the growing significance of the Cerebellum as an area of increasing research interest in the context of cognitive impairment.

\section*{Material and methods}
Our analysis approach involves three main steps: (1) Creating a FCN for each subject in each group, (2) Estimating a group representative matrix using the self-attention deep model, and (3) Extracting group representative features of ROIs connections using LSIRM and visualizing them on the group summary FCN.

\subsection*{ADNI study}

The ADNI dataset is composed of four consecutive cohorts (ADNI1, ADNI2, ADNI--GO, and ADNI3). Participants were recruited for initial periods in the ADNI1 cohorts (October 2004). Follow--up of participants were recruited to the ADNI3 cohort period. To facilitate preprocessing of the fMRI data, we filtered data with the same acquisition protocols as the database. A total of three protocol conditions (200 timepoints, TR = 3000 ms, 48 slices) were applied for selection. After filtering based on three conditions, a total of 281 participants remained in the ADNI2, ADNI-go, and ADNI3 cohorts. The ADNI1 cohort was excluded because it did not contain data that met the aforementioned conditions. As a result, we used axial rs-fMRI data from 57 AD subjects, 93 EMCI subjects, 53 LMCI subjects, and 78 MCI subjects (Fig.~\hyperref[fig:proposal]{1a}). By focusing on these specific disease pairs, we aim to uncover the key ROIs that exhibit distinct patterns and contribute significantly to the classification and differentiation of these cognitive impairment conditions. All data are publicly available, at ~\url{http://adni.loni.usc.edu/}.

\subsection*{MRI acquisition}

The participants included in this study participated in scanning at diverse sites through 3T MRI scanners manufactured by Philips Medical Systems or Siemens Healthineers. The detailed MRI protocols of the ADNI dataset were reported in the webpage (\url{http://adni.loni.usc.edu/methods/mri-tool/mri-acquisition/}). In the ADNI2 and ADNI-go cohorts, MRI scanning was performed at twenty-six different sites with Philips 3T MRI scanners, using synchronized scanning parameters. In the case of the ADNI3 cohort, Siemens 3T MRI scanners were used to collect fMRI data with synchronized parameters.

\subsection*{MRI preprocessing} 

The ADNI database's scanned imaging data underwent a thorough quality check by trained analysts. This process consisted of two stages of quality control. The first stage involved examining the consistency of protocol parameters, while the second stage focused on checking series-specific quality factors such as body motion, anatomical coverage, and other potential artifacts. After these two stages, each image was assigned one of four quality labels (1 to 3 indicating acceptable levels and 4 indicating unusable). We processed resting-state fMRI (rs-fMRI) data that met our acceptability criteria for research purposes.
To extract time courses from regions of interest (ROIs) from rs-fMRI data, we utilized SPM12 (\url{www.fil.ion.ucl.ac.uk/spm/}) and the DPARSFA toolbox (V5.1, \url{http://rfmri.org/dpabi}). The standard preprocessing pipeline for ROI time course extraction was employed, which included slice-time correction, realignment, normalization with an Echo Planar Imaging(EPI) template, detrending and smoothing with a 6mm kernel.
Temporal filtering within a range from 0.01Hz to 0.1Hz was carried out to eliminate physiological noises. After preprocessing steps were completed, we obtained temporal signals.

\subsection*{Functional connectivity networks (FCNs)}\label{step1}
Dimension Reductions methods are the well-known method to embed the complex structure data such as Principal Component Analysis (PCA)~\cite{dunteman1989principal}, T-stochastic neighbor embedding (t--SNE)~\cite{van2008visualizing}, and Uniform Manifold Approximation Projection (UMAP)~\cite{mcinnes2018umap}. t-SNE and UMAP model the manifold using stochastic (i.e., converting neighborhood’s distance into conditional probability that represents similarity) and topological (i.e., fuzzy simplicial complex with edge weights representing the likelihood of connectivity) information, respectively.

\paragraph{Dimension reduction} 
PCA~\cite{dunteman1989principal} is a technique that uses orthogonal transformation to reduce high-dimensional data to low-dimensional data. It converts high-dimensional space samples that are likely to be related to each other into low-dimensional space samples (main components) that are not linearly related. The axis with the most significant variance is the first principal component, and the second greatest variance is the second principal component. This decomposition divides the sample into the components that best represent the differences of information that have important implications for data analysis. On the other hand, t-SNE~\cite{van2008visualizing} is a non-linear dimension reduction method that aids in understanding the data with impact information. It is based on t-distribution, which is comparable to normal distribution but has the heavy tail component that is helpful in covering up the far distribution element of high--dimensional data. When two data construct similar structures, they nearly correspond to each other based on the similarity value from the t-distribution. The t-SNE results depict the embedded points whose distances, trained by calculating the points' resemblance in structure. reflect their degree of similarity. UMAP~\cite{mcinnes2018umap} is a nonlinear dimension reduction method that models the manifold using a topological structure. Because it is based on topological space, the embedding points are close in proximity if the two data points have similar topological features. It first reorganizes the data into a fuzzy simplicial complex, which then produces the connections based on the hyper-parameter that controls the connectivity around the data. Then, it projects the correlated structured data into a low-dimensional space based on their connection, where the connection indicates the aforementioned close proximity.

\paragraph{Mapper}
Mapper is a one of techniques derived from topological data analysis, which allows us to represent the topological structure of high-dimensional data as a network. Topological data analysis simplifies the complexity of the topological space by transforming it into a network consisting of nodes and connections that capture the topological characteristics, such as points, lines, and triangles, within the data. The Mapper process involves two main steps. First, the high-dimensional topological space is mapped onto a measure space, typically a real space, represented as a graph. This mapping function can be any real-valued function that captures the essential features of the data. In the next step, the mapper partitions the graph into subsets of data, and clustering is performed within each subset. This process explores the interrelationships between the subsets, identifying the structural relationships within the data. The result of this process is called the Mapper, where each cluster becomes a node, and nodes are connected when they share similar data attributes.

The Mapper can be considered as a form of partial clustering. It applies a standard clustering algorithm to subsets of the original data and examines the interactions between the resulting sub-clusters. When two non-empty subsets U and V are considered, their sub-clusters may have overlapping elements, which are used to construct a simplicial complex. The sub-clusters themselves are referred to as vertices or nodes, while the overlapping elements form edges in the complex. This process yields a simplicial complex consisting of dots, lines, and triangles, which provides insights into the topological structure of high-dimensional data.

\subsection*{Attention distribution matrix}
\label{step2}

Given that the FCN represents a correlated network of connections between ROIs, the self--attention deep learning model is suitable for handling FCN data~\cite{velickovic2017graph,lei2022longitudinal,zhang2022pairwise}.

\paragraph{Self--Attention deep learning model}
Attention mechanism is used to focus on specific input values from sequence--based tasks that is most relevant to the input in order to reduce information loss and increase information power~\cite{bahdanau2014neural}. The attention mechanism utilizes attention scores to emphasize the important factors during model training. These scores are determined by the query, key, and value which are three components in attention mechanism. These elements are often represented as vectors. First, the query vector conveys information about the current element calculated by the attention module. On the other hand, the key vector contains context or information associated with each element in the sequence. These keys are used to compute attention scores by measuring the similarity between the query and each key. The value vector holds the content associated with each element. Once the attention scores are calculated, they are employed to assign weights to the values. The resulting weighted sum of these values corresponds to the attended information or context that holds relevance to the present query. In short, the attention distribution comprises both keys and queries. In this context, the value serves the purpose of modeling the attention distribution in relation to the output. Self-attention, also known as intra-attention, refers to an attention mechanism that establishes connections between various positions within a single sequence, aiming to generate a comprehensive representation of that sequence.

Since our input data is the adjacency matrix among ROIs calculated from the FCN, it is better to train the interactions among ROIs. To model this interaction, self-attention mechanism is suggested. Self-attention mechanism allows the model to relate different elements of the input sequence to each other, capturing dependencies and relationships between elements. In summary, for a specific ROI represented as a query, we compute its similarity (attention distribution) with all other ROIs (represented as keys) and then take a weighted sum based on these attention scores. This process helps the model learn which ROIs are highly relevant to the task at hand, such as disease classification. In other words, the value of key and query is equal. Ultimately, by utilizing the learned attention distribution, we can examine the relationships between different ROIs. 

An attention mechanism can be conceptualized as a process that maps a query and a collection of key-value pairs to generate an output. In this scenario, all components - query, keys, values, and output - are represented as vectors. The output is determined through a weighted summation of the values, and these weights are calculated by a compatibility function that takes into account the relationship between the query and the corresponding key. The input is made up of queries and keys, both of which have a dimensionality of $d_k$, and values that have a dimensionality of $d_v$. The computation involves calculating dot products between the query and all keys, followed by division of each result by the square root of $d_k$, and subsequently applying a softmax function to acquire weights corresponding to the values. If we consider a total of $\textbf{R}$ ROIs, we can represent $\mathbf{Q} \in \mathbb{R}^{\textbf{R} \times \textbf{R}}$, $\mathbf{K} \in \mathbb{R}^{\textbf{R} \times \textbf{R}}$, and $\mathbf{V} \in \mathbb{R}^{\textbf{R} \times \textbf{R}}$. The attention distirbution matrix $attn(\mathbf{Q},\mathbf{K})$ is then calculated as: 

\begin{equation}
\begin{split}
 attn(\mathbf{Q},\mathbf{K}) = \text{Softmax}(\frac{\mathbf{Q}\mathbf{K}^{\top}}{\sqrt{d_k}}) \\
 \text{Attention}(\mathbf{Q},\mathbf{K},\mathbf{V}) = attn(\mathbf{Q},\mathbf{K})\mathbf{V} 
\end{split}
\end{equation}

%If we assume there are a total \textbf{R} number of ROIs, we can express that $\mathbf{Q} \in \mathbb{R}^{\textbf{R} \times \textbf{R}}, \mathbf{K} \in \mathbb{R}^{\textbf{R} \times \textbf{R}}$ and $\mathbf{V} \in \mathbb{R}^{\textbf{R} \times \textbf{R}}$. The input consists of queries and keys of dimension $d_k$, and values of dimension $d_v$. We compute the matrix of output as:

% Binary classification full model. 
% 어떤 파트의 어떤 process를 통해 attention distribution이 세워지는지 

To account for various aspects of ROIs, as single-head attention might not adequately capture the information, especially in high-dimensional datasets, the equation is extended to multi-head attention. In this expansion, $H$ sets of weight layers, denoted as $W_i^Q$, $W_i^K$, and $W_i^{V}$, are applied to each output layer of the attention model, specifically to $\mathbf{Q}$, $\mathbf{K}$, and $\mathbf{V}$. The multi-head attention is created by concatenating these $H$ sets of $head_i$, where $i=1\cdots H$.
\begin{equation}
\begin{split}
 \text{MultiHead}(\mathbf{Q},\mathbf{K},\mathbf{V}) = \text{Concat}(\text{head}_{1},...,{\text{head}}_{H})W^O \\
\text{where }head_i=\text{Attention}(\mathbf{Q}W^{\mathbf{Q}}_{i},\mathbf{K}W^{\mathbf{K}}_{i},\mathbf{V}W^{\mathbf{V}}_{i})
\end{split}
\end{equation}

Note that $W_i^Q \in \mathbb{R}^{d_\text{model} \times d_k}$, $W_i^K \in \mathbb{R}^{d_\text{model} \times d_k}$, $W_i^V \in \mathbb{R}^{d_\text{model} \times d_v}$ and $W^O \in \mathbb{R}^{{hd_v} \times d_{\text{model}}}$. 

For each individual input, the attention distribution matrix  $attnM(\mathbf{Q},\mathbf{K})$ can be derived by averaging the attention layers across all $H$ layers, denoted as $attnH(\mathbf{Q},\mathbf{K})_i$.

\begin{equation}
\begin{split}
attnM(\mathbf{Q},\mathbf{K}) = \frac{\sum_{i=1}^H{attnH(\mathbf{Q},\mathbf{K})_i}}{H} \\
\text{where }attnH(\mathbf{Q},\mathbf{K})_i= attn(\mathbf{Q}W^{\mathbf{Q}}_{i},\mathbf{K}W^{\mathbf{K}}_{i})
\end{split}
\end{equation}

%%%%% JJKIM %%%%%
% 

\subsection*{Group representative ROIs features }\label{step3}

Treating the group representative matrix $\mathbf{X}{h|g,h}$ as an item-response dataset, each item $j$ corresponds to an ROI, and the responses $i$ represent the subjects. The LSIRM enables us to identify common ROIs that exhibit similar patterns across subjects within each group.

Through the LSIRM, we estimate the latent positions for the ROIs, which are determined based on the relationships among the subjects. If a particular ROI demonstrates a consistent pattern among subjects, its latent position will be located near the origin, as indicated by close distances between the ROI and other subjects. Consequently, by analyzing the latent space item-response model, we can identify the ROIs that generally demonstrate consistent responses across subjects within each group.

Furthermore, we can compare the latent positions of ROIs between corresponding pairs of groups. For instance, if a specific ROI's latent position is near the origin in Group $g$ but located outside the origin in Group $h$, we can infer that this ROI exhibits a distinguishing feature in Group $g$, indicating a different pattern compared to Group $h$. The LSIRM thus facilitates the identification of ROIs that play a role in distinguishing between the two pairs of disease groups.

\paragraph{Latent Space Item-Response Model (LSIRM)}
LSIRM~\cite{jeon2021mapping} is a model that treats item-response structure datasets as bipartite networks and estimates the interactions between items and respondents. In our study, we aim to estimate the latent positions of ROIs based on the interactions between subjects and ROIs. While the original LSIRM model is designed for item-response datasets~\cite{embretson2013item} where each cell value is either 0 or 1, we adapt continuouse version of LSIRM to group representative matrix $\mathbf{X}{h|g,h}$ where each cell value is continuous. This adaptation allows us to effectively model the relationships between ROIs and sujbects in our specific context. Equation \eqref{eq:lsrm} shows the continuous version of LSIRM: 
\begin{equation}\label{eq:lsrm}
\mathbb{P}(y_{ij}\mid \boldsymbol{\Theta})  \sim \text{Normal} ( \theta_j +\beta_i - || {\bf u}_j - {\bf v}_i ||, \sigma^2 ) .
\end{equation}
where $y_{ij}$ is coefficient of variation of ROI $j$ in attention distribution of subject $i$, $i=1,\cdot,N_g$ , and $j=1,\cdots, R$. Each $\boldsymbol{\Theta}$  represents  $\{\boldsymbol{\theta}=\{\theta_j\},\boldsymbol{\beta}=\{\beta_i\}, \bf{U}= \{{\bf u}_j\}, \bf{V}= \{{\bf v }_i\} \}$ and $|| {\bf u}_j - {\bf v}_i ||$ represents the Euclidean distance between subject $i$ and ROI $j$. LSIRM consists of two parts: the attribute part and the interaction part. In the attribute part, there are two parameters: $\theta_j \in \mathbf{R}$ and $\beta_i \in \mathbf{R}$. The parameter $\beta_i$ represents the number of ROIs that have non-zero values for subject $i$, while $\theta_j$ represents the number of subjects that react (have non-zero values) to ROI $j$. In the interaction part, we have the latent configurations ${\bf u}_j$ and ${\bf v}_i$ for each ROI $j$ and subject $i$, respectively. These latent positions allow us to estimate the interactions between subjects and ROIs. Specifically, the latent positions are estimated based on the distances between the latent positions of other subjects and ROIs. By examining the estimated latent positions of the ROIs, we can identify which ROIs are commonly reacted to among subjects. For example, if the latent position ${\bf u}_j$ of an ROI is estimated to be near the origin, it indicates that most subjects show similar patterns of reaction to that ROI. This property can be utilized to extract commonly reacted ROIs from each group $h$ using the group representative matrix $\mathbf{X}_{h|g,h}$. \\

To estimate parameters in LSIRM, we use Bayesian inference. We specify prior distribution for the parameters:
\begin{equation}
\begin{split}
    \beta_i \lvert \tau^2_\beta &\sim \text{N}(0,\tau^2_\beta),\quad \tau^2_\beta >0 \\
    \theta_j \lvert \sigma^2 &\sim \text{N}(0,\sigma_{\theta}^2),\quad \sigma^2 >0 \\
    \sigma^2 &\sim \text{Inv-Gamma}(a , b),\quad a_>0 , \quad b>0 \\
    \sigma_{\theta}^2 &\sim \text{Inv-Gamma}(a_\sigma , b_\sigma),\quad a_\sigma>0 , \quad b_\sigma>0 \\
    \bf{u}_j &\sim \text{MVN}_d (\mathbf{0, I}_d) \\
    \bf{v}_i &\sim \text{MVN}_d (\mathbf{0, I}_d).
\end{split}
\end{equation}
where $\mathbf{0}$ is a $d$--vector of zeros and $\mathbf{I}_d$is the $d \times d$ identify matrix. The posterior distribution of LSIRM is proportional to
\begin{equation}
\begin{split}
    \pi(\boldsymbol{\Theta}, \sigma^2 \mid \bf{Y} ) &\propto \prod_{j}\prod_{i} \mathbb{P} \left (y_{ji} \mid \boldsymbol{\Theta} \right)^{y_{ji}} \left( 1-\mathbb{P} \left (y_{ji} \mid \boldsymbol{\Theta} \right)\right)^{1-y_{ji}} \\
    &\times \prod_j \pi(\theta_{j}\mid \sigma_{\theta}^2 ) \pi(\sigma_{\theta}^2) \prod_i \pi(\beta_i)  \\
    &\times \prod_j \pi(\bf{u}_{j} ) \prod_i \pi( \bf{v}_i)  \pi(\sigma^2)
\end{split}
\end{equation}
\section*{Data availability}

The datasets analysed during the current study are available in the Alzheimer’s Disease Neuroimaging Initiative (ADNI) repository, ~\url{https://adni.loni.usc.edu/}.

\bibliography{sample}

\section*{Acknowledgements}

This research was supported in part by the Brain Research Program through the National Research Foundation of Korea(NRF) funded by the Ministry of Science and ICT (2017M3C7A1029485) and in part by the National Research Foundation of Korea (NRF) Grant through the Korea Government [Ministry of Science and ICT (MSIT)] under Grant 2019R1A2C1007399.

\section*{Author information}

These authors contributed equally: Jeong-Jae Kim, Yeseul Jeon.

\subsection*{Authors and Affiliations}

\noindent \textbf{Graduate Program in Cognitive Science, Yonsei University, Seoul, Republic of Korea}

\noindent Jeong-Jae Kim, Junggu Choi \& Sanghoon Han

\noindent \textbf{Department of Statistics and Data Science, Yonsei University, Seoul, Republic of Korea}

\noindent Yeseul Jeon

\noindent \textbf{Department of Psychology and Neuroscience, Duke University, NC, USA}

\noindent SuMin Yu

\noindent \textbf{Department of Psychology, Yonsei University, Seoul, Republic of Korea}

\noindent Sanghoon Han

\subsection*{Contributions}

Conceptualization: J.J.K., Y.J., and S.H.; methodology: J.J.K., Y.J., and S.H.; validation: S.H.; formal analysis: J.J.K. and Y.J.; investigation: J.J.K., Y.J., S.Y., J.C., and S.H.; writing--original draft preparation: J.J.K. and Y.J.; writing--review and editing: J.J.K., Y.J., S.Y., J.C., and S.H.; rs-fMRI preprocessing: J.C.; visualization: J.J.K. and Y.J.; supervision: S.H.; project administration: S.H. All authors have read and agreed to the submitted version of the manuscript.

\subsection*{Corresponding author} 

Correspondence to \href{mailto:sanghoon.han@yonsei.ac.kr}{Sanghoon Han}.

\section*{Ethics declarations}

\subsection*{Competing interests}

The authors declare no competing interests.

\section*{Supplementary information}

\paragraph{\href{https://drive.google.com/file/d/1f_FSjT5QLzmA_CmZcgRD5M5MZJs-KTRM/view?usp=sharing}{Supplementary Information}}
\label{supp}

%\section*{Source Data}

%\href{https://drive.google.com/file/d/1fqRjPH1tmk16dY6lIt7VAEx6dyeqjrrG/view?usp=sharing}{Source Data File}

\end{document}